\documentclass[review]{elsarticle}
\usepackage{amsmath}
\usepackage{makecell}
\usepackage{lineno,hyperref}
\usepackage{multirow}
\usepackage{array}
\usepackage{rotating}
\usepackage{graphicx}
\usepackage{booktabs}
\usepackage{ctable}
\usepackage{tabularx}
\usepackage{color}
\makeatletter

 \newcommand*{\@rowstyle}{}
\newcommand*{\rowstyle}[1]{
  \gdef\@rowstyle{#1}%
  \leavevmode\@rowstyle
  \ignorespaces
}
\newcolumntype{=}{
  >{\gdef\@rowstyle{}\ignorespaces}%
}
\newcolumntype{+}{
  >{\leavevmode\@rowstyle\ignorespaces}%
}

 \newcolumntype{C}[1]{>{\centering\arraybackslash}p{#1}}
 \makeatother

\modulolinenumbers[2]
\usepackage[a4paper, total={6in, 8in}]{geometry}
\journal{Applied Energy}

\begin{document}

\begin{frontmatter}

\title{Application of Information Gap Decision Theory in Practical Energy Problems: A Comprehensive Review }

\author[tabriz]{M. Majidi}
\author[tabriz]{B. Mohammadi-Ivatloo}
\author[UCD]{A. Soroudi\corref{cor1}}
\cortext[cor1]{\textcolor[rgb]{0,0,0} {Corresponding author: School of Electrical, Electronic and Communications Engineering, University College Dublin, Dublin, Ireland }}
\ead{majidmajidi95@ms.tabrizu.ac.ir}
\ead{bmohammadi@tabrizu.ac.ir}
\ead{alireza.soroudi@ucd.ie}

\address[tabriz]{Smart Energy Systems Laboratory, Faculty of Electrical and Computer Engineering, University of Tabriz, Tabriz, Iran}
\address[UCD]{University College Dublin, Dublin, Ireland}

\begin{abstract}
The uncertainty and risk modeling is a hot topic in  operation and planning of energy systems. The system operators and planners are decision makers that need to handle the uncertainty of input data of their models. 
As an example, energy consumption has always been a critical problem for operators since the forecasted values, and the actual consumption is never expected to be the same.
The penetration of renewable energy resources is continuously increasing in recent and upcoming years. These technologies are not dispatch-able and are highly dependent on natural resources. This would make the real-time energy balancing more complicated. Another source of uncertainty is related to energy market prices which are determined by the market participants' behaviors.  
To consider these issues, the uncertainty modeling should be performed. Various approaches have been previously utilized to model the uncertainty of these parameters such as probabilistic approaches, possibilistic approaches, hybrid possibilistic–probabilistic approach, information gap decision theory, robust and interval optimization techniques. This paper reviews the research works that used information gap decision theory for uncertainty modeling in energy and power systems.
\end{abstract}

\begin{keyword}
Uncertainty\sep uncertain parameters \sep information gap decision theory\sep robustness function\sep opportunity function \sep energy.
\end{keyword}

\end{frontmatter}

\section{Introduction} \label{Introduction}
\begin{figure}[ht]
	\centering
	\includegraphics[width=9.5cm, height=7.5cm]{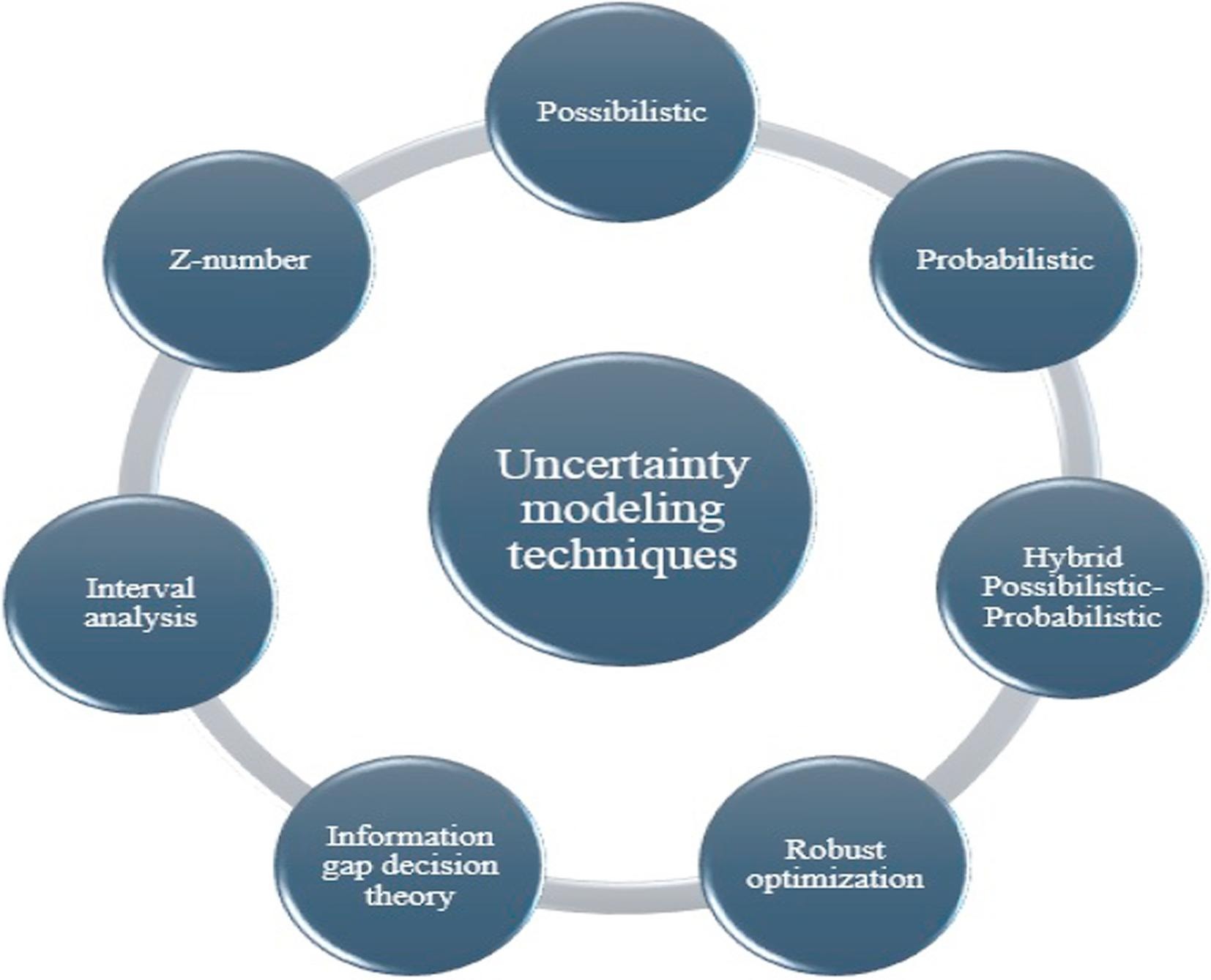}
	\caption{Techniques and approaches used for uncertainty modeling \textcolor[rgb]{0,0,0} {\cite{soroudi2013decision}}}
	\label{fig:1}
\end{figure}

The decision making and optimization is a part of almost every engineering problem \cite{soroudi2017power}. The energy systems is not an exception in this regard. The optimal decisions depend on the problem structure as well as the accuracy of input data. 
The uncertainty of different input parameters in \textcolor[rgb]{0,0,0} {energy} systems has made several operating issues for the system operators and other stakeholders in this area. 
The uncertainty influences the scheduled plans and might create new challenges for the involved decision makers. To resolve the issues mentioned above, various methods and techniques have been studied and utilized to control the outcomes caused by possible uncertainty in the behavior of parameters. Previously implemented techniques according to \cite{soroudi2013decision} are possibilistic technique \cite{gouveia2017dc}, probabilistic technique \cite{dantzig2004linear}, hybrid possibilistic–probabilistic techniques \cite{soroudi2012possibilistic}, robust optimization technique \cite{soyster1973convex,nazari2018robust}, information gap decision theory (IGDT) \textcolor[rgb]{0,0,0} { \cite{ben2006info,jabari2019introduction}} and interval analysis \cite{moore2009introduction}. The mentioned approaches and techniques are briefly illustrated in Fig. \ref{fig:1}. Each approach has specific characteristics and uses different ways to model uncertainty which will be explained in the next section.

As mentioned before, the primary goal of this study is to investigate the \textcolor[rgb]{0,0,0} {research works} that used information gap decision theory in \textcolor[rgb]{0,0,0} {energy system} studies and reveal unexplored areas. The information gap decision theory models the positive and negative aspects of uncertainty based on the known and unknown information. Positive and negative outcomes that uncertainty may \textcolor[rgb]{0,0,0} {cause} are modeled using two functions of information gap decision theory called robustness and opportunity functions. These functions will be comprehensively explained in the next sections.
The rest of the proposed paper can be expressed as follows: Uncertainty modeling techniques and approaches are briefly explained in Section \ref{sec:UMAEIGDT}. Information gap decision theory and corresponding functions are explained in Section \ref{sec:IGDT}. 
Papers in the field of \textcolor[rgb]{0,0,0} {energy} system studies in which information gap decision theory has been employed are investigated and reviewed in Section \ref{sec:RBIGDT}. Finally, the conclusions are presented in Section \ref{sec:conclusion}.

\section{Uncertainty modeling approaches excluding IGDT}
\label{sec:UMAEIGDT}
The challenges caused by the realization of different uncertain parameters in the \textcolor[rgb]{0,0,0} {energy} system has encouraged operators to use uncertainty modeling techniques to become ready against possible consequences and make the best decision. Uncertain parameters can be numerous and should be investigated from economic and technical points of view. The well-known ones to be named are energy demand, possible outages of components, the output of renewable generation units and market price. The methods used for modeling mentioned uncertainties are briefly described in this section \cite{soroudi2013decision}.
\subsection{Possibilistic technique}
\textcolor[rgb]{0,0,0} {In this technique, the parameters showing uncertain behavior are considered to be $X$}. The system model is assumed to be $f$ and the output variable is considered to be $y$. To define the membership function of $y$ while the membership function of $X$ is known, \textcolor[rgb]{0,0,0} {the} $\alpha$-cut approach is used \cite{janssen2013monte}. Finally, using a defuzzification technique such as centroid technique, max-membership principal, weighted average approach, the fuzzy members are converted to crisp members.
 \subsection{Probabilistic technique }
In this method, a multivariate function \textcolor[rgb]{0,0,0} {like} $y=f(x)$ \textcolor[rgb]{0,0,0} {is in hand}. Different random parameters are set to be inputs to the system, $x$, \textcolor[rgb]{0,0,0} {for} which \textcolor[rgb]{0,0,0} {the} probability density function is known. $y$ is considered to be the output with an unknown probability density function, and the $f$ is assumed to be the system model. Totally, there are three major probabilistic approaches modeling uncertainty which are: scenario-based technique \cite{morales2009scenario}, point estimate technique \cite{hong1998efficient} and Monte Carlo simulation \cite{kalos2008monte}.
 \subsection{Hybrid possibilistic-probabilistic techniques}
In the cases that both of possibilistic and probabilistic uncertain parameters are available, hybrid possibilistic–probabilistic techniques are employed to model uncertainty. These techniques are possibilistic-scenario based technique \cite{soroudi2012possibilistic} and possibilistic-Monte Carlo technique \cite{soroudi2011possibilistic}.
\subsection{Robust optimization technique}
In some cases, the uncertainty is modeled as an uncertainty set. It means that the uncertain parameters always belong to a known uncertainty set. In these cases, the robust optimization technique can be a good a choice to be used for uncertainty modeling. In this approach, the worst possible condition is determined and then using the obtained results, appropriate strategies are \textcolor[rgb]{0,0,0} {followed} \cite{ben2009robust}. The degree of being conservative is adjustable by the decision maker. 
\subsection{ Interval analysis}
In this method, the range of uncertain parameters is \textcolor[rgb]{0,0,0} {known}, and then the upper and lower bounds of the objective function are obtained as the main goal of interval problem \cite{moore2009introduction}. 
\subsection{Z-number}
\textcolor[rgb]{0,0,0} {This method assesses how reliable the info is. It includes two components: The first one is a limitation for real-valued uncertain variable and the second one is a measure of uncertainty for the first one \cite{zadeh2011note}}.

\section{Information gap decision theory (IGDT)}
\label{sec:IGDT}
\begin{figure}[ht]
	\centering
	\includegraphics[width=0.3\columnwidth,trim=1cm 0.8cm 0.4cm 0.91cm, angle = 0]{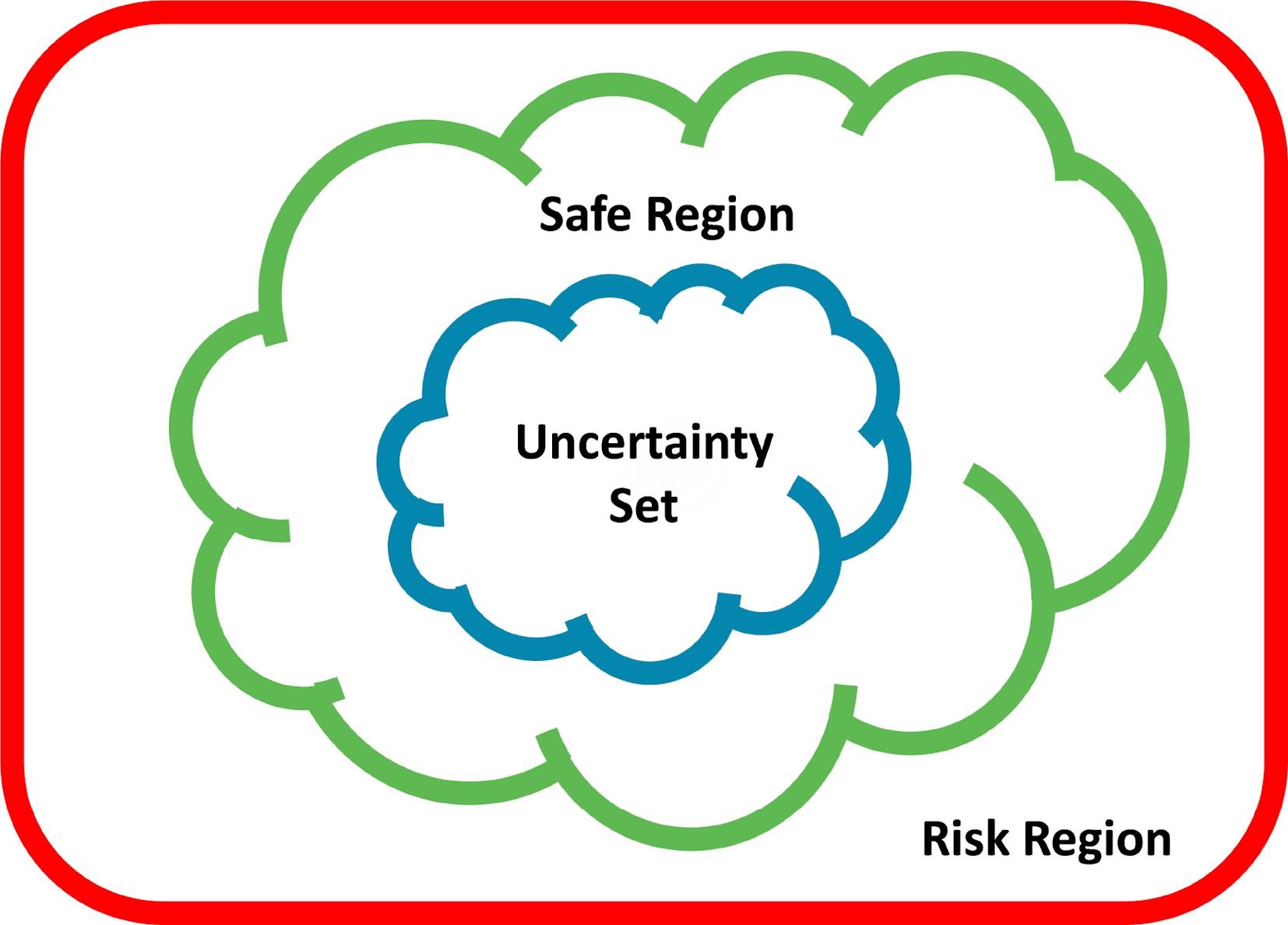}
	\caption{Safe/risk region of uncertainty based problem}
	\label{fig:2}
\end{figure}

\begin{figure}[ht]
	\centering
	\includegraphics[width=0.3\columnwidth,trim=1cm 0.8cm 0.5cm 0.91cm, angle = 0]{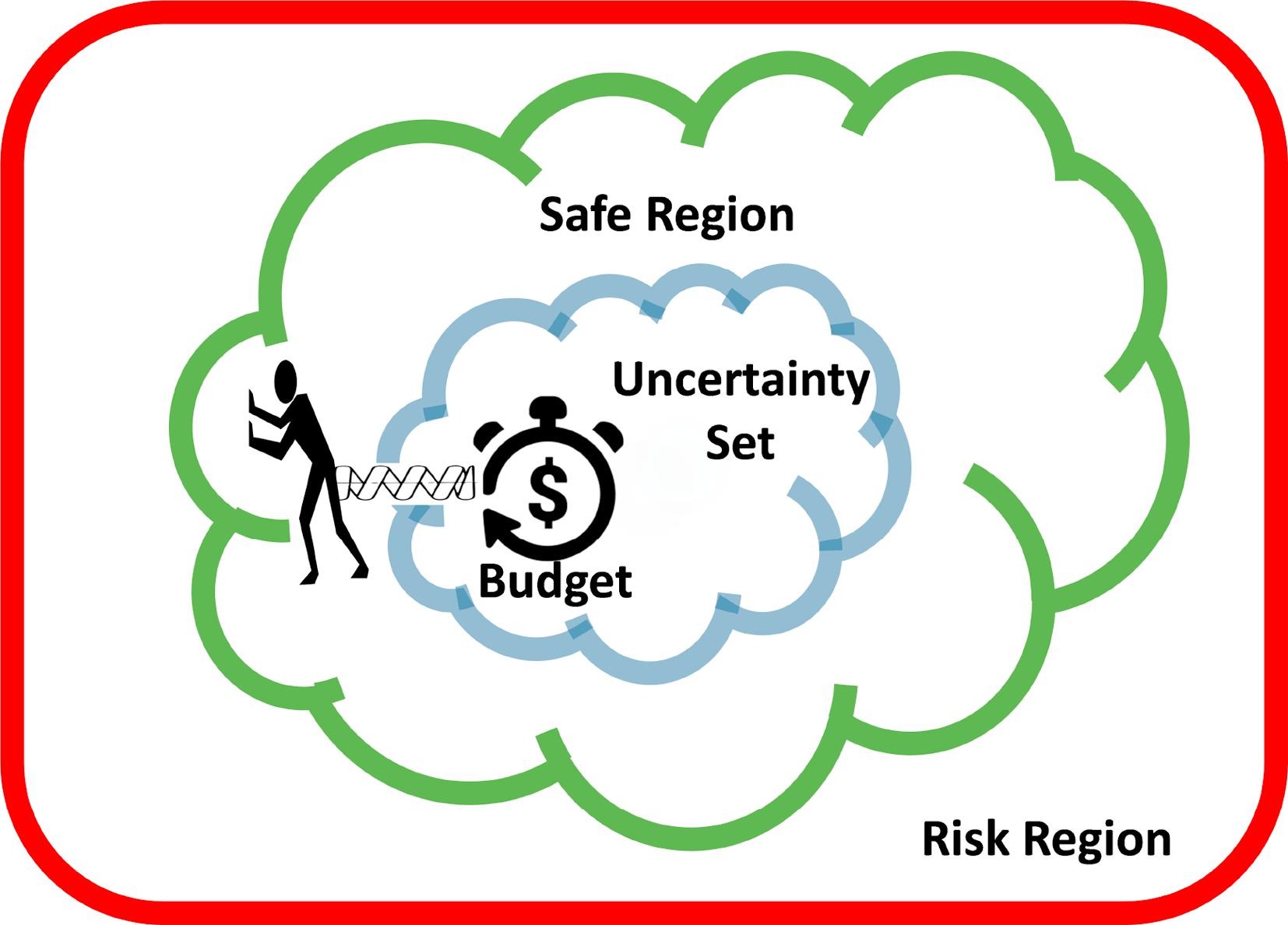}
	\caption{Graphical explanation of IGDT}
	\label{fig:3}
\end{figure}

In this section, \textcolor[rgb]{0,0,0} {IGDT} is described in general to give a better and clear view of IGDT application in \textcolor[rgb]{0,0,0} {energy} system scheduling and planning.
At first, some of the papers in various fields except \textcolor[rgb]{0,0,0} {energy} system scheduling that employed IGDT for uncertainty modeling are summarized.\\
\textcolor[rgb]{0,0,0} {IGDT is employed for modeling uncertainty of biological value persistence or presence in reserve sites in \cite{moilanen2006planning}, investigating model updating in \cite{ben2001info}, risk-based designing of civil structure in \cite{takewaki2005info}, modeling uncertain of behavior of a system including unknown events over time in \cite{ben2009info}, solving a spatial search-planning problem including unclear probabilistic info and data in \cite{sisso2010info}, modeling uncertainties in ecosystems in \cite{hildebrandt2011investment}, modeling uncertainties in neural networks in \cite{pierce2006evaluation}, studying environmentally benign manufacturing and design in cite{duncan2008including} and controlling uncertainties in water resources in \cite{hipel1999decision}}. \\
\indent \textcolor[rgb]{0,0,0} {Additionally, IGDT is used for obtaining the highest possible social welfare under various circumstances in \cite{stranlund2008price}, formulating structural design-codes in \cite{ben1999design}, practical applications in \cite{ben2002graph}, solving the risk-based portfolio management problem in finance in \cite{berleant2008portfolio} and obtaining appropriate uncertainty-based sites for CO2 sequestration in \cite{grasinger2016decision}. 
It should be noted that set-models of IGDT are discussed in \cite{ben1999set} and information gap models are constructed about probability distributions using probability bounds analysis and p-boxes in \cite{ferson2008probability}}.\\
\indent Unlike other uncertainty modeling approaches, IGDT does not need a \textcolor[rgb]{0,0,0} {huge} amount of data for uncertainty modeling. By exploiting the available data about uncertainty parameter, IGDT informs the operator about the negative and positive results that can be caused by uncertainty to take appropriate and logical decisions which may be safe or risky. \textcolor[rgb]{0,0,0} {IGDT receives the uncertain uncertainty set which may not be exact. Afterward, IGDT tries to make the system performance \textcolor[rgb]{0,0,0} {immune} against the uncertain parameter while keeping the operational point within the safe region. Sometimes, the uncertainty level of uncertain parameter is such severe that system may not withstand the possible instabilities caused by the mentioned uncertainty. This is one of features of IGDT to assure that system does not get into the risky region}.\\ 
\indent Fig. \ref{fig:2} is used to illustrate the safe/risky region of a problem experiencing uncertainty to which IGDT is due to be applied. Sometimes when uncertainty has occurred, operators need to take strategies to handle the condition caused by uncertainty. IGDT is an efficient tool to evaluate and compare the strategies taken at the times of uncertainty, and the decision maker would be able to evaluate the effectiveness of each strategy, determine his priorities and evaluate his expected objective function \cite{ben2006info}. \textcolor[rgb]{0,0,0} {The uncertainty levels determined by IGDT in the safe region are not always free enough to be extended as much as possible. There are usually different factors that may influence the resistance level of system against uncertainty. IGDT tries to maximize the resistance level of operating system against uncertainty within the safe region while satisfying the other factors that may limit the extension of safe uncertainty set.}
For example, consider a system participating in energy market for its energy demand satisfaction. If the market price is uncertain \cite{MEHDIZADEH2018617}, the system operator should increase its operating budget to be robust against the possible increase in price. \\ \indent \textcolor[rgb]{0,0,0} {The budget limitation is the factor that may affect the taken policies against uncertainty. So, for a reasonable amount of increase in budget, the operator expects to obtain the maximum degree of robustness against the market price increase. This IGDT-based problem is a bi-level optimization problem in which the uncertainty set should be maximized and the operational \textcolor[rgb]{0,0,0} {cost} should be minimized. The mentioned example is schematically illustrated in Fig. \ref{fig:3}. According to this \textcolor[rgb]{0,0,0} {figure}, the uncertainty set (the level of system resistance) is tried to be enhanced as much as possible while the operational budget for controlling the uncertainty is limited.}

\vspace{4cm}

\subsection{ Immunity functions of IGDT}
\begin{figure}[ht]
	\centering
	\includegraphics[width=12.7cm, height=4.2cm]{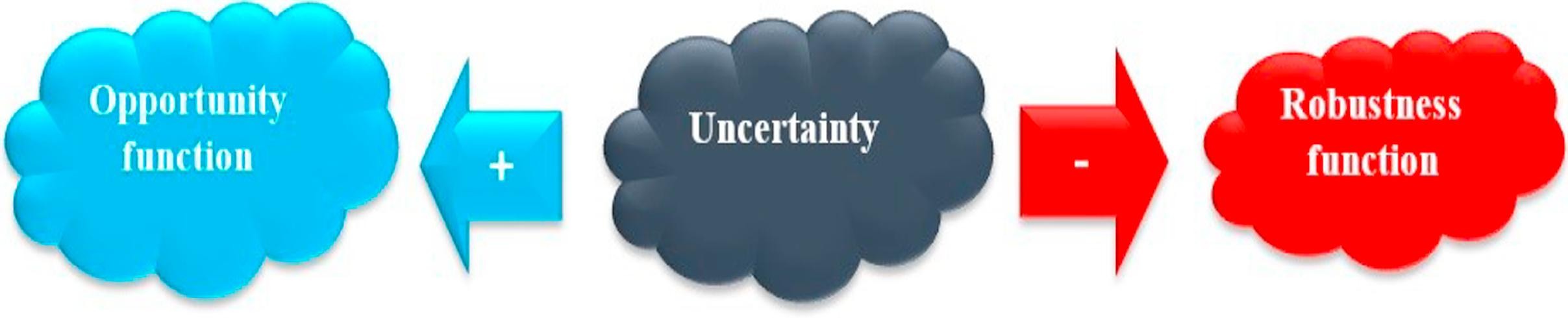}
	\caption{The robustness and opportunity functions of IGDT}
	\label{fig:4}
\end{figure}

Information gap decision theory includes two immunity functions namely robustness and opportunity functions which are depicted in Fig.    \ref{fig:4}. Immunity functions can be used by relevant decision makers to model positive and negative outcomes of uncertainty. These functions are known as major and significant parts \textcolor[rgb]{0,0,0} {of} IGDT since \textcolor[rgb]{0,0,0} {operator's} decisions are based on these functions. Predicting possible results and outcomes of uncertainty, robustness and opportunity functions facilitate evaluation of various conditions and help operators take the best strategies \textcolor[rgb]{0,0,0} {when} uncertainty \textcolor[rgb]{0,0,0} {occurs} \cite{ben1999set, ferson2008probability}. \textcolor[rgb]{0,0,0} {As it can be seen from Fig.    \ref{fig:4}, robustness function of IGDT is usually employed to model negative impacts of uncertainty toward which system should be resistant. Alternatively, as shown with positive sign in Fig. \ref{fig:4}, the opportunity function is used to assess the possible benefits obtainable from uncertainty.} 

\subsubsection{Robustness function}
In order to address \textcolor[rgb]{0,0,0} {the} negative aspects \textcolor[rgb]{0,0,0} {of} uncertainty, \textcolor[rgb]{0,0,0} {the} robustness function is used. $\alpha$ \textcolor[rgb]{0,0,0} {is the parameter that models this function}. The greatest level of uncertainty that can be handled through the taken strategies is determined by this function \cite{ben2006info}:
\begin{equation}
\hat \alpha  = \mathop {\max }\limits_\alpha  \left\{ {\alpha :\,\,maximum\,\,total\,\,cost\,\,is\,\,not\,higher\,\,than\,\,a\,\,specified\,\,limit\,} \right\}
\end{equation}
It can be concluded from \textcolor[rgb]{0,0,0} {(1)} that a greater value of \textcolor[rgb]{0,0,0} {$\hat \alpha$} is \textcolor[rgb]{0,0,0} {more desirable}.
\subsubsection{Opportunity function}
Benefits and positive outcomes that can be achieved through uncertainty are modeled by opportunity function \cite{ben2006info}:
\begin{equation}
\hat \beta  = \mathop {\min }\limits_\beta  \left\{ {\alpha :\,\,\,minimum\,\,total\,\,cost\,\,is\,\,less\,\,than\,\,a\,\,specified\,\,limit\,} \right\}
\end{equation}
It can be concluded from \textcolor[rgb]{0,0,0} {(2)} that a smaller value of \textcolor[rgb]{0,0,0} {$\hat \beta$} is \textcolor[rgb]{0,0,0} {more desirable}.
 \subsection{Structure of IGDT}
Information gap decision theory is mainly composed of three parts: system model, operation requirements and uncertainty modeling.
\subsubsection{System model}
The relationship between inputs and outputs is expressed by system model. Based on IGDT, system model can be expressed by $y (q, l)$ in which $q$ is decision variable and $l$ is uncertainty parameter  \cite{ben2006info}. For example, in an economic problem with the objective function cost and market price as the uncertain parameter, system model can be expressed by $C (q, l)$ in which $C (q, l)$ is the cost function as the system model, $q$ is decision variable and l is market price. IGDT uses uncertainty set to strengthen \textcolor[rgb]{0,0,0} {the} objective function against the uncertainty of input parameters.

 \subsubsection{Operation requirements}
Information gap decision theory includes two main functions namely robustness function
 $\hat{\alpha}(q)  $   and opportunity function, $  \hat{\beta}(q)  $  . Each \textcolor[rgb]{0,0,0} {function} is used to simulate positive and negative aspects of uncertainty, and then according to the achieved results, the appropriate decisions are taken by the operator of the system. The robustness function expresses how resistant the system is against the increase of uncertain parameter. In simple words, robustness functions determine how much more money should be paid to avoid further harmful outcomes.
\textcolor[rgb]{0,0,0} {Let's} consider the same economic problem with market price uncertainty. Based on the robustness function, the maximum resistance is desired while the total cost of the system should not exceed a predefined value. Mathematical formulation of this example is expressed by \textcolor[rgb]{0,0,0} {equation (3)} \cite{ben2006info}.
\begin{equation}
\hat{\alpha}(C_{r})=max_\alpha\{\alpha: max(C(q,l))\le C_{r}\}
\end{equation}
$ C_{r} $ is the defined cost that \textcolor[rgb]{0,0,0} {the} maximum total cost of system cannot exceed.
On the other hand, \textcolor[rgb]{0,0,0} {the} opportunity function of information gap decision theory is used to determine how the system can benefit from the possible reduction of the uncertain parameter which is the positive effects of uncertainty. In the same example, a possible reduction of market price can bring economic benefits to the system. \textcolor[rgb]{0,0,0} {In this case}, \textcolor[rgb]{0,0,0} {the} opportunity function is used which mathematical formulation is presented in the following \cite{ben2006info}:
\begin{equation}
\hat{\beta}(C_{o})=min_\beta\{\alpha: min(C(q,l))\le C_{o}\}  
\end{equation}
$ C_{o} $ is the defined cost that minimum total cost of the system cannot \textcolor[rgb]{0,0,0} {exceed it}.
Its noteworthy that in \textcolor[rgb]{0,0,0} {(4)}, $ C_{o} $   is lower than  $ C_{r} $.
 \subsubsection{Uncertainty model}
\begin{figure}[ht]
	\centering
	\includegraphics[width=8cm, height=3.8cm]{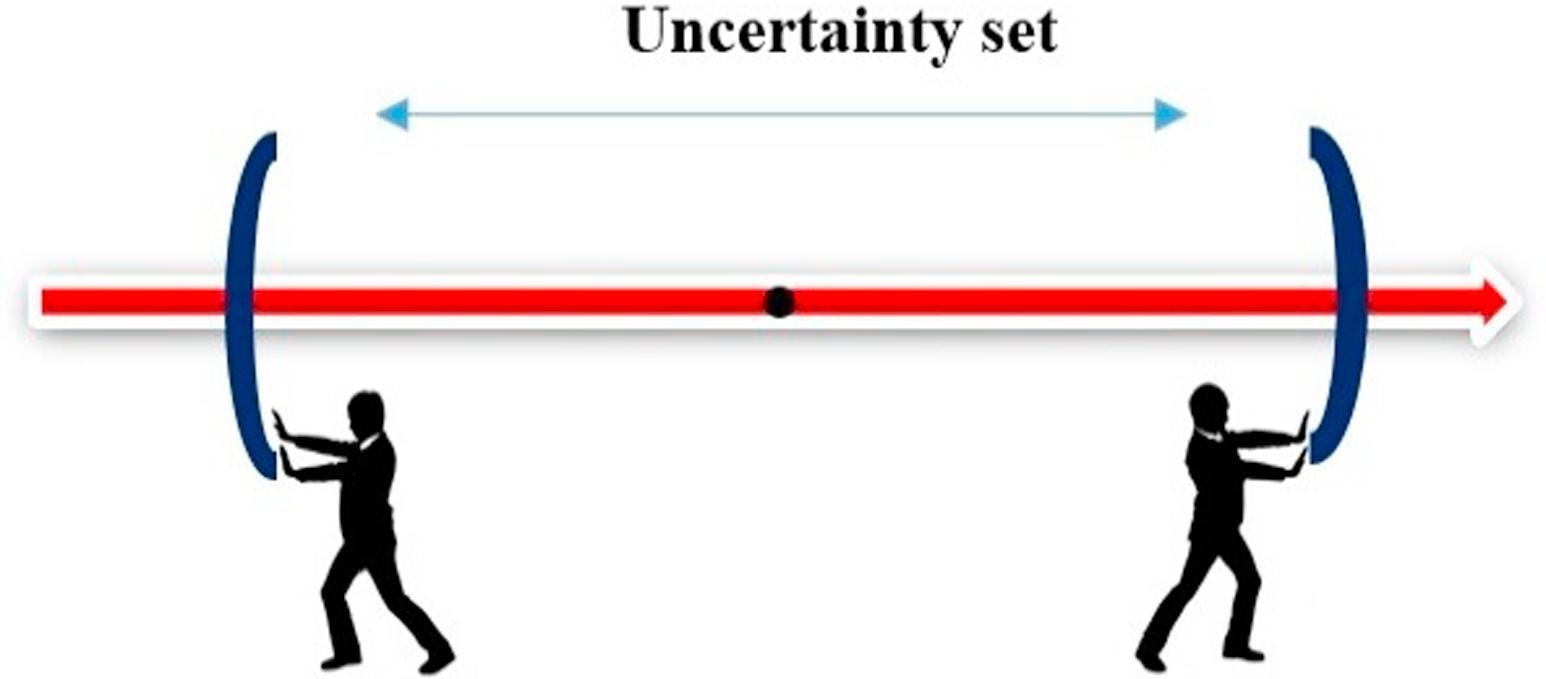}
	\caption{Uncertainty model of envelope bound of IGDT }
	\label{fig:5}
\end{figure}

\textcolor[rgb]{0,0,0} {Various} info-gap models are available for uncertainty including envelope-bound models, energy-bound models, Minkowski-norm models, Slope-bound models, Fourier-bound models, Hybrid info-gap models, combined info-gap models. The whole uncertainty models along with their relevant mathematical formulation are presented in table \ref{tab1} \cite{ben2006info}.\\
\textcolor[rgb]{0,0,0} {As it can be observed, different uncertainty models with specific features and formulas are available. One of the most popular models (approved by analyzing the employed models in literature gathered in Table 2) is envelope bound model which is illustrated in Fig.    \ref{fig:5} \cite{7546906}. According to the uncertainty model of envelope bound model presented in this figure, the uncertainty set in IGDT method is tried to be maximized. By having the uncertainty set maximized, the most possible resistance level can be obtained for the operating system against uncertainty. Also, under such a condition, the most possible benefits can be achieved while the requirements are satisfied.} 

\begin{table}[!t]
 \caption{Uncertainty models existing in IGDT literature}
	\makebox[\linewidth]{
\centering
{\footnotesize

\scalebox{0.9}{
	\begin{tabular}{ | p{1cm} | p{1.2cm} | p{8.5cm} | p{5.5cm} | }
		\hline
    	\multicolumn{2}{ |c| }{Model} & Mathematical  form & Description \\ \hline
	   \multicolumn{2}{|p{2.5cm}|}{Envelope-bound models}  &   $ U(\alpha , \tilde{u})=\{u(t):| \dfrac{u(t)-\tilde{u}(t)}{\varphi(t)} | \le \alpha  \} , \alpha \ge 0 $ & $u(t)$ is uncertain parameter and $ \tilde{u}(t) $ is the nominal value of  uncertain   parameter \\ \hline
		\vspace{1mm} \multirow{2}{1cm}{Energy-bound models} &    Scalar  functions & $ U(\alpha , \tilde{u})=\{u(t):\int_{0}^{\infty} [{u(t)-\tilde{u}(t)}]^{2}dt \le \alpha^{2}  \} , \alpha \ge 0 $  &$ u(t) $ is uncertain scalar function and $ \tilde{u}(t) $  is the nominal function  \\ \cline{2-4}
		& Vector  functions  &  $ U(\alpha , \tilde{u})=\{u(t):\int_{0}^{\infty} [{u(t)-\tilde{u}(t)}]^{r} V[u(t)-\tilde{u}(t)] dt \le \alpha^{2} \} , \alpha \ge 0 $  & $ u(t) $  is uncertain vector function,  $ \tilde{u}(t) $is the nominal vector function and $ V $ is known positive definite  \\
		\hline
		\multicolumn{2}{|p{2.5cm}|}{Minkpwski-norm models}  & $ U_{r}(\alpha , \tilde{u})=\{u:||V^{\dfrac{1}{2}}(u-\tilde{u}||_{r}\le \alpha)\} , \alpha \ge 0$ &  $ u  $is uncertain parameter and $ \tilde{u} $  is nominal value of uncertain parameter  \\
			\hline
			\multicolumn{2}{|p{2.5cm}|}{Slope-bound models}  &   $ U(\alpha , \tilde{u})=\{u(t):| \dfrac{du(t)-\tilde{u}(t)}{dt} | \le \alpha\varPsi(t)  \} , \alpha \ge 0 $ & $ u  $  is uncertain vector function, $ \tilde{u} $  is the nominal vector function and $ \varPsi $  determines the envelope of uncertain variation of the slope \\
			\hline
		 \multicolumn{2}{|p{2.5cm}|}{Fourier-bound models }  & $ U(\alpha , \tilde{u})=\{u(y)=\tilde{u}+c^{T}Wc\le\alpha^{2}\}, \alpha \ge 0 $ & $ u $  is uncertain vector function, $ \tilde{u} $ is the nominal vector function and  $ W $  is a known , symmetric, positive matrix \\
			\hline
	        	\vspace{1mm} \multirow{2}{1.1cm}{Hybrid info-gap models} &    Continues random variable & $ U(\alpha , \tilde{p})=\{p(x):p(x)\in \rho, |p(x)-\tilde{p}(x)| \le \alpha \varPsi(t) \}, \alpha \ge 0 $  &$ x $  is random variable, $ p(x) $  is probability density function of $ x $  and $ \tilde{p} $  is the best known of
 $ p(x) $
	        	\\ \cline{2-4}
	        & Discredit random variable  & $ U(\alpha , \tilde{p})=\{p:p\in \rho , ( p-\tilde{p}^{T}V (p-\tilde{p}) \le \alpha^{2} \}, \alpha \ge 0 $  & Mp  is probability distribution of random variable and $ \tilde{p} $ is the  best known prediction of the probability distribution of random variable  \\
	        \hline
	
	        \multicolumn{2}{|p{2.5cm}|}{Combined info-gap models}  & $ U(\alpha , \tilde{u})= \{u(t) : |u(t)-\tilde{u}|\le \alpha \varPsi(t) , \int_{0}^{\infty}[u(t)-\tilde{u}]^{2} dt \le \alpha^{2}  \} , \alpha \ge 0 $ &  $ u  $is uncertain parameter and $ \tilde{u} $  is nominal value of uncertain parameter  \\
	        \hline
	
	         \multicolumn{2}{|p{2.5cm}|}{Discrete info-gap models}  & $ \Pi^{k}=\{\pi\pm e^{i} , \forall \pi \in \Pi^{k-1} , i=1,...,J \} , k=1,2,... $&  $ J $  is number of options, $ \pi $  is preference vector and $ \Pi^{k} $  is set of preference vectors differ from the nominal by no more than $ k $  single preference changes  \\
	        \hline

	        	      	\end{tabular}
}
}
}

\label{tab1}
\end{table}

\section{ Risk-based energy problems solved by IGDT}
\label{sec:RBIGDT}
In this section, the cases and \textcolor[rgb]{0,0,0} {energy} problems to which IGDT has been applied are classified and summarized. For more clarification, the cases are classified into three groups \textcolor[rgb]{0,0,0} {according to the type of uncertain parameter: 1) market price 2) generation/demand 3) both.}\\
\textcolor[rgb]{0,0,0} {One of the appropriate solutions to deal with the uncertain behavior of input data in scheduling and planning of energy systems is using of energy storage systems. Various technologies of storage systems have been employed in energy systems to mitigate the negative impact of uncertainty. As an example, CAES has been utilized to cope with stochastic nature of uncertainties like wind \cite{aliasghari2018look,akbari2019stochastic} and demand \cite{jadidbonab2019cvar}. Additionally, pumped hydro storage (PHS) has been used to handle the uncertainties like the uncertainties of stream flow \cite{kocaman2017value}, solar systems \cite{sun2019optimal}, wind units \cite{yildiran2018risk,golshani2018coordination,xia2019milp} and loads \cite{simab2018multi}}.
\subsection{ Risk-based problems with the uncertainty of market price }
\textcolor[rgb]{0,0,0} {The market price uncertainty is one of essential issues in energy systems that might affect the decisions of involved stakeholders. Different sectors like retailers, generation companies and storage systems in different energy problems need to deal with this issue to guarantee a safe region for operating the system.}
\begin{itemize}
      \item Retailer\\ 
 \indent In a competitive energy market, consumers with low energy demands do not procure their energy demands from energy market. Instead, the retailers do participate in the market on behalf of them. Retailers purchase energy at a variable price form pool market and sell it to the consumers at a fixed price. The retailer attempts to consider possible uncertainties of market price to have a robust performance. This uncertainty modeling is done by IGDT in \cite{charwand2014midterm}. \\ 
 \indent \textcolor[rgb]{0,0,0} {A retailer who purchases energy from different sources and then sells it to the end-users should make a good balance between the purchased power and the power due to be consumed to gain the expected profit. Taking into account this issue, a novel reliability-based optimization framework has been developed in \cite{khojasteh2018reliability} in which IGDT and two-point estimate methods have been employed to model the uncertainties of wholesale price and rival’s price, respectively}. The robust participation of retailers owning distributed generation units in energy markets \textcolor[rgb]{0,0,0} {is} investigated in \cite{khojasteh2015decision} in which IGDT has been employed to obtain risk-averse and risk-taking strategies in energy markets. The uncertainty of prices at which retailer procures energy from pool market is modeled using IGDT in \cite{kazemi2015risk}.
    \item Large consumers\\  
 \indent Large consumers can participate in energy markets to meet its energy demand. Bilateral contracts and pool markets are available options for large consumers to supply its demand. Employing robustness and opportunity functions, various energy procurement strategies have been provided for large consumers to take appropriate decisions while participating in energy market \cite{zare2010multi}.\\ 
 \indent It should be noted that in addition to the energy market, distributed generation units have also been used to supply some parts of energy demand. In a day-ahead market with uncertain hourly prices, uncertainty modeling of price is vital. Therefore, Variance-covariance matrix of IGDT has been utilized in \cite{zare2010demand} to obtain robust strategies for the participation of large consumer in the energy market. \textcolor[rgb]{0,0,0} {Similar problem is studied with envelope-bound model of IGDT under demand response in \cite{nojavan2016robust}}. Efficient utilization of energy carriers can result in higher efficiency up to 80 \% for \textcolor[rgb]{0,0,0} {energy} systems. In \cite{alipour2016optimal}, large consumers with co-generation systems like CHP systems have participated in the energy market to supply energy demands under uncertainty of energy market price that is modeled using IGDT.
     \item Unit commitment\\ 
\indent \textcolor[rgb]{0,0,0} {Unit commitment problem of thermal energy units has been evaluated through an IGDT-constrained optimization platform with considering environmental issues in \cite{jabari2019risk,soroudi2017information}.}
    \item Genco\\
\indent  With the deregulation in the energy sector, generation companies are seeking to maximize their profits through optimal incorporating distributed generation units. Produced power by these units is sold in power market and therefore, uncertainty modeling of market price is done by IGDT in \cite{mohammadi2013application}. Also, similar to the problem solved in \cite{kaviani2016milp}, a bi-level optimization problem is investigated in \cite{kazemi2013igdt} in which bilateral contracts are taken into account by generation company to reduce the risk of participation in energy market as much as possible. The market price uncertainty is modeled by IGDT in \cite{kazemi2013igdt}.\\ 
\indent Generation companies have different generation resources which can be either renewable or non-renewable. In the case of non-renewable ones, different types of fuels are burned which leads to environment pollutants.The  generation company participates in three electricity, emission and fuel markets in which price of electricity, emission and fuel are considered to be uncertain. These uncertainties are modeled using IGDT in \cite{mathuria2014igdt}. The robust performance of a thermal generation company in day-ahead and subsequent adjustment markets has been investigated in \cite{mathuria2016robust}. In \cite{kazemi2015risk}, risk-based bi-level problem of a utility has been investigated. Gained profit due to participation of utility in wholesale market plus the income gained from contracted demand of retail side of utility results the total profit of utility. Using information gap decision theory, a competition based biding problem has been solved for generation company in \cite{cheong2004information}.  Electricity utilities can join energy markets to supply energy demand requested by related retailers. The total benefit of mentioned utilities is provided through participation in the electricity market and selling the procured energy to the relevant retailers.
To maximize the utilities benefit, the uncertainty of market price should be taken into account to avoid utility facing unstable conditions. This problem has been investigated in the presence of \textcolor[rgb]{0,0,0} {responsive and non-responsive} loads in \cite{kazemi2014risk} to determine appropriate participating strategies to maximize utilities total benefit.  
    \item Storage\\
\indent As a distributed generation unit, compressed air energy storage system (CAES) can participate in energy market to presents power in various time periods and therefore gain benefit. CAES attempts to forecast market price so an uncertainty modeling is necessary. Therefore, IGDT based uncertainty modeling of market price is done for CAES in \cite{shafiee2017risk}.
    \item Microgrid\\
\indent \textcolor[rgb]{0,0,0} {As a small scale energy system, microgrid can have an essential role in generating local energy and satisfying energy consumer's demands. A novel IGDT-based optimization model is developed for a microgrid in accordance with game theoretical approach in  \cite{dehghan2018optimal} to optimize the energy provision of the microgrid operator subject to uncertainty of wholesale prices. A hub energy based microgrid system equipped to different energy facilities has been optimally scheduled under different uncertainties in  \cite{dolatabadi2019short}. In detail, the uncertainty of demand and wind forecasts have been modeled through a scenario-based method while the uncertainty of electricity price within the mentioned system has been modeled using IGDT.}\\   
\indent \textcolor[rgb]{0,0,0} {As one of important energy sectors, smart homes equipped to smart facilities have been optimally scheduled under uncertainties within the research papers. These systems can rely on smart energy equipment to optimally satisfy the end-users' requirements while controlling the possible uncertainties. In detail,  \cite{najafi2018information} has developed an optimization model according to which apartment smart buildings capable of purchasing power from different sectors like market has been optimally scheduled and IGDT concept is employed to assess risk-involved performance of apartment smart building toward market price uncertainty. The impact of thermal storage has been investigated through having different case studies. The smart home equipped to smart energy technologies has been optimally configured in \cite{najafi2018robust} under uncertainty of market price. To handle the mentioned uncertainty for efficiently using the available energies, robustness and opportunity functions of IGDT have been used.}    
\end{itemize}
 
\subsection{Risk-based problems with the uncertainty of generation/consumption }

As mentioned in former sections, different parameters in energy systems can have uncertain behavior. In addition to the market price, the load and generation can also have uncertain performance. In this section, risk-based performance of energy systems with uncertain load or generation has been reviewed. The main focus is on renewable resources like wind, photo-voltaic and etc. Due to the variability of wind speed, the output of wind turbine fluctuates and affects the generation-load balance of the system. In order to solve such problems, the uncertainty modeling is necessary.

\begin{itemize}

      \item Microgrid\\ 
\indent \textcolor[rgb]{0,0,0} {Efficient utilization of clean energies like solar has turned to be an essential topic in the research papers. An optimization model based on IGDT is proposed in \cite{nojavan2019risk} to optimize uncertainty-based performance of a solar power plant equipped with thermal storage.\\ 
\indent Load uncertainty of power systems has been always one of the major challenges that system operators have been faced. Various load forecasting methods and techniques are available but, in order to take the forecasting tolerances into account, uncertainty modeling should be done. In \cite{nojavan2017performance}, uncertainty based performance of an on-grid hybrid energy system with electrical and thermal energy demands has been investigated. Electrical load of mentioned hybrid system is considered to be uncertain. Using robustness and opportunity functions of IGDT, uncertainty modeling is done and appropriate power procurement strategies are obtained to be used by system operator. In order to improve economic performance and reduce total operation cost of hybrid system, the same problem has been investigated in \cite{nojavan2017risk} in the presence of demand response program. In \cite{jabari2018risk}, an industrial continuous heat treatment furnace based on air source heat pump is optimally scheduled to supply heating demand under extreme temperatures. IGDT method is employed to model the uncertainty of heating demand and provide risk-averse and opprotunity-seeking strategies.} 
      \item Transmission \& Distribution Sector\\ 
 \indent  \textcolor[rgb]{0,0,0} {To consider the uncertainties of both wind units and PV systems, information gap decision theory is employed in \cite{hooshmand2018robust} in which demand response and storage models have been developed to enhance the system performance within the mentioned uncertainties. In a similar research in \cite{rabiee2018information}, wind-based energy system has been planned under wind unit’s uncertainty. Uncertainty of wind units has been modeled by IGDT in long term subject to security constraint like voltage stability index to assure safe performance of system. Smart hub energy systems in distribution networks have been optimally scheduled subject to technical imitations like power flow constraint under different uncertainties in \cite{majidi2019integration}. To model the uncertainties of upstream network price, renewable units and electrical demands of smart hubs, interval optimization, scenario-based method and IGDT are respectively employed.}\\
 \indent  Because of large scale renewable generation, wind farms with several wind-turbines should be prepared for each ramp event of wind turbines. Wind-turbine ramp events can make network unstable and therefore, prediction of such mentioned events are vital. using information gap decision theory, uncertainty of wind power ramp prediction has been modeled in \cite{ma2015igdt}. Optimal power flow problem in the presence of wind generation uncertainty has been evaluated in \cite{murphy2016information}. Also, optimal power flow problem under uncertainty offshore wind farms has been investigated in \cite{rabiee2015information} where HVDC technology has been used to connect onshore AC network to the offshore wind farms. In addition to the power generation for direct consumption of consumers in demand sides, generation units allocate some percentage of their generation capacity for reserve generation. Depending on that participation in reserve market is beneficial are not, generation units injects reserve power to the transmission network. Uncertainty of presented reserve by generation units to the transmission line is investigated through IGDT in \cite{zhao2016flexible}. Finally, uncertainty modeling of wind generation in a power system with DC model is done in \cite{maghouli2016robust} in which energy storage system has been employed to mitigate disturbing effects caused by uncertainty of wind generation.\\
 \indent  Load uncertainty problem is investigated in \cite{sarhadi2015robust} in which risk-based expansion planning of transmission line has been evaluated under uncertain behavior of load. \textcolor[rgb]{0,0,0} {In \cite{rezaei2018energy}, a novel optimization framework has been presented for uncertainty-based energy management of an islanded microgrid in which IGDT methodology has been implemented to consider the uncertainties of renewable unit’s generation and energy demand}. In \cite{li2015robust}, uncertainty modeling of transmission line overload has been assessed using IGDT in which robust performance of power system has been obtained for the cases line experiences overload. Sometimes, distribution systems may experience some outages which are harmful for structure of the whole system. In thus condition, restoration strategies become more important. Factors influencing restoration strategies are uncertainty of some parameters in power systems which should be molded using uncertainty modeling approaches. In \cite{chen2015robust}, uncertainty of load and output power of distributed generation units have been modeled using IGDT to improve restoration results of power system in the cases of outages. \textcolor[rgb]{0,0,0} {Similarly, an IGDT-based optimization model has been presented for load restoration in \cite{xie2019second} in which the uncertainty of load has been modeled with IGDT to ensure the maximum tolerable load increment}. Finally, the transmission network reinforcement using series reactance is investigated with considering uncertainty of wind generation in \cite{soroudi2017resiliency} in which uncertainty modeling is carried out by IGDT.

      \item Unit Commitment\\ 
\indent \textcolor[rgb]{0,0,0} {Security-involved unit commitment problem has been solved subject to uncertainty of load in \cite{ahmadi2019security} in which novel storage model has been developed and IGDT methodology is employed to model the load uncertainty. In another study, linear relationships have been presented in \cite{razavi2018robust} to model linear form of unit commitment problem in which the uncertainty demand has been taken into account with IGDT. In \cite{nikzad2019robust}, unit commitment problem has been solved subject to uncertainties of demand and renewable unit’s generation via robust optimization method and IGDT and the results of employed methodologies are presented to derive the necessary conclusions. Unit commitment problem has been also solved for integrated electricity and gas networks in \cite{mirzaei2019igdt} in which the impact of uncertainty of wind generation on performance of operating system has been assessed with IGDT}. Similarly, the unit commitment problem has been investigated without considering security constraints in \cite{soroudi2017information} where demand response program has been used to improve economic performance of system.\\  
\indent \textcolor[rgb]{0,0,0} {Taking into account power flow constraints in the scheduling of energy sectors can make the results more real and accurate So, a novel optimization model based-on mathematical bases has been presented in \cite{nazari2019network} to optimize performance of an integrated heat and power energy sector under different uncertainties. In detail, the studied system is equipped to different energy resources like combined heat and power systems, boiler units and other resources and to consider the uncertainties of market price, renewable units and electric demand, respectively, robust optimization, scenario-based method and IGDT are employed. The results depicted that implementation of mentioned methods can help the operator to take appropriate decisions against uncertainties.} Finally, \textcolor[rgb]{0,0,0} {with taking into account the uncertainty of responsive loads participation in demand response programs (considered as reserve providers), IGDT-based optimization framework has been developed in  \cite{ghahary2018optimal} in which IGDT methodology has been employed to provide the decisions-making strategies}. 

      \item Retailer\\ 
\indent \textcolor[rgb]{0,0,0} {In \cite{golmohamadi2018bi}, to solve the electricity pricing and dispatch issues that retailer of local generation unit may deal with, a novel uncertainty-based optimization framework based on bi-level programming has been developed in which the scenario-based method and IGDT have been respectively employed in the upper level and lower level problems to model the related uncertainties.}

\end{itemize}
 
\subsection{Risk-based problems with the uncertainty of price and generation/consumption }

Problems with generation/load and price as the uncertain parameter were investigated in previous sections. Here, the problem with both price and generation/load uncertainties are reviewed. 

\begin{itemize}
      \item Transmission \& Distribution system\\ 
\indent In \cite{shivaie2016risk} a risk-based multi-objective optimization model has been presented for generation and transmission planning problems in which electrical load in total investment cost of the system have been considered to be uncertain parameters for which uncertainty modeling has been done using IGDT. Using the melody search algorithm and Powell heuristic approach, non-convex transmission expansion planning problem is solved in \cite{shivaie2016strategic} in which electrical load and market price are both considered to be uncertain parameters of the system.\\ 
\indent In a deregulated power system, distribution network operator (DNO) supplies the demanded energy by consumers through available energy resources. In addition to the types of energy resources, DNO is due to choose, the uncertainty of electrical demand as well as the uncertainty of market price are challenging for DNO \cite{soroudi2013igdt}. In \cite{mazidi2016design}, in addition to the uncertainty of electrical load and market price, uncertainty modeling of wind generation is also considered via IGDT, and consequently, optimal decision-making strategies are obtained for DNO to take the best possible decisions.  
      \item Retailer\\ 
\indent The retailer can either participate in the energy market to purchased energy, or he can use local distributed generation units to generate power locally. There exist two challenges for the retailer in both cases: in the first case, the uncertainty of market price and in the second case, the uncertainty of local units output can make problems for the retailer. So, to handle such mentioned problems, retailer attempts to model uncertainty of market price as well as local units output to have robust performance in the energy market and maximize his benefit \cite{xue2016application}.  \textcolor[rgb]{0,0,0} {Uncertainty-based participation of demand response aggregator in demand response programs has been investigated through an IGDT-based optimization framework in \cite{vahid2019self} in which the uncertainties of electricity price in market as well as participation level of consumers in demand response have been modeled via IGDT.} 
      \item GenCo\\ 
\indent In former sections, output uncertainty problem of renewable generation units like the wind was completely discussed. In studied papers, the only generation of the wind turbine was considered to be the uncertain parameter for which uncertainty modeling was done. In \cite{moradi2015self}, in addition to generation, owner of the wind turbine is faced with the uncertainty of market price. Therefore, using information gap decision theory, uncertainty modeling has been for both outputs of the wind turbine and market price to be robust against possible fluctuation of each mentioned uncertain parameters.
\end{itemize}
\subsection{Summary of reviewed papers}
For more clarification, studied papers in the field of IGDT are classified according to various bases in table \ref{tab:summary}. \textcolor[rgb]{0,0,0} {Analyzing this table, useful info about the uncertain parameter, type of uncertainty model and type of programming can be derived.}
\begin{table}[htp]\color{blue}
\caption{Summary of reviewed papers in the field of IGDT}		
\label{tab:summary}
{\footnotesize
	\makebox[\linewidth]{
\begin{tabular}{|p{2.2cm}|c|c|c|c|c|c|c|c|}
			\hline
	{\footnotesize  paper}	& \multicolumn{3}{c|}{Uncertain parameter} &{\footnotesize Uncertainty model} &{\footnotesize  Programming}   & \multicolumn{3}{c|}{  {\footnotesize  Planning period}}\\
			\cline{2-4} \cline{7-9}
			& \textbf{Price} & \textbf{Power$^{1}$}  & \textbf{Both}  &   & &\textbf{SH-T}&\textbf{M-T}&\textbf{L-T}\\

			\hline
			\cite{zare2010electricity,zare2011risk,zare2010demand,nojavan2016robust,kazemi2014risk,khojasteh2015decision,alipour2016optimal,khojasteh2018reliability,shafiee2017risk,dehghan2018optimal,dolatabadi2019short,cheong2004information,najafi2018robust} & * &  &  & Envelope-bound & MINLP  & * &  & \\
			\hline
			\cite{zare2010multi}& * &  &  & Variance-covariance matrices & MINLP&  * &  &  \\ 
			\hline
            \cite{charwand2014midterm}& * &  &  & Envelope-bound & MINLP&  & * &  \\
			\hline
			\cite{nojavan2015hybrid,mohammadi2013application,kazemi2013igdt,kazemi2013igdt,kazemi2015risk,jabari2019risk}& * &  &  & Envelope-bound  & MINLP  & * &  &  \\
			\hline
			\cite{nojavan2013risk}& * &  &  & Variance-covariance matrices & MINLP  & * &  &  \\
			\hline
			\cite{kaviani2016milp,mazidi2016incorporating}& * &  &  & Envelope-bound  & MILP &  * &  &  \\
			\hline
			\cite{mathuria2016robust}& * &  &  & Fourier-bound models  & MINLP  & * &  &  \\
			\hline
			\cite{najafi2018information,najafi2018heating}& * &  &  & Envelope-bound  & MILP &  * &  &  \\
			\hline
			\cite{nojavan2019risk,soroudi2017resiliency,rezaei2018energy,golmohamadi2018bi}&  & * &  & Envelope-bound  & MILP  & * &  &  \\
			\hline
			\cite{hooshmand2018robust,murphy2016information,zhao2016flexible,li2015robust,chen2015robust}&  & * &  & Envelope-bound  & MINLP & * &  &  \\
			\hline
			\cite{ahmadi2019security,razavi2018robust}&  & * &  & Envelope-bound  & MILP& * &  &  \\
			\hline
			\cite{nikzad2019robust,mirzaei2019igdt,soroudi2017information,ma2015igdt,ghahary2018optimal}&  & * &  &  Envelope-bound  & MINLP &  * &  &  \\
			\hline
			\cite{rabiee2018information}&  & * &  & Envelope-bound  & NLP &  &  & * \\
			\hline
			\cite{nazari2019network,majidi2019integration}&  & * &  & Envelope-bound  & MINLP &  * &  &  \\
			\hline
			\cite{rabiee2015information}&  & * &  & Envelope-bound  & NLP  & * &  & \\
			\hline
			\cite{maghouli2016robust}&  & * &  & Envelope-bound  & MIQCP&  &  & * \\
			\hline
			\cite{nojavan2017performance,nojavan2017risk}&  & * &  & Envelope-bound   & MIP &   &  & * \\
			\hline
			\cite{sarhadi2015robust}&  & * &  & Envelope-bound  & MINLP &  &  & * \\
			\hline
			\cite{xie2019second}&  & * &  & Envelope-bound  & SOCP & * &  &  \\
			\hline
			\cite{jabari2018risk}&  & * &  & Envelope-bound  & NLP  & * &  &  \\
			\hline
			\cite{dehghan2014multi}&  &  & * & Envelope-bound  & MILP  &  &  & *  \\
			\hline
			\cite{shivaie2016risk,shivaie2016strategic}&  &  & * & Envelope-bound  & MINLP  &  & *& \\
			\hline
			\cite{moradi2015self,xue2016application,soroudi2013igdt,vahid2019self}&  &  & * & Envelope-bound   & MINLP & * &  &  \\
			\hline
            \cite{mazidi2016design}&  &  & * & Envelope-bound  & MILP & * &  & \\
            \hline
         \multicolumn{3}{c}{1-Generation/Consumption}
		\end{tabular}
}

}

\label{tab2}

\end{table} 

\newpage

\section{Conclusion}
\label{sec:conclusion}
The uncertainty as one of the major challenging issues for system operators has been broadly studied by many researchers. Since some parameters in \textcolor[rgb]{0,0,0} {energy} systems have uncertain nature and according to the fact that the forecasts are not always exact, therefore uncertainty modeling is a vital issue in \textcolor[rgb]{0,0,0} {energy} system studies. As investigated in this paper, many different risk modeling techniques are available that can be used for uncertainty modeling in \textcolor[rgb]{0,0,0} {energy} systems with various structures and conditions. Each risk modeling techniques has its unique features that differentiate these techniques from one another. It can be concluded that the approach that needs less info in compared with other methods is \textcolor[rgb]{0,0,0} {information gap decision theory}. The \textcolor[rgb]{0,0,0} {information gap decision theory} involves two main immunity functions that are robustness and opportunity functions. Robustness functions are used to model robustness degree of system faced with uncertainty against the possible increase of risk.
\textcolor[rgb]{0,0,0} {Alternatively, the} opportunity function can be used to determine how benefits can be achieved from possible 'good behaving' of the uncertain parameter. 

The untouched research areas are uncertainty modeling of  participation of demand response as well as plug-in hybrid electric vehicles aggregators in energy markets, risk-based designing and planning problems of multi-carrier energy systems and  micro-grids, security-based designing problem of substations, load growth uncertainty modeling, uncertainty modeling of power plants as well as line availability, uncertainty based designing problem of transmission towers under different weather conditions, disturbance uncertainty modeling in transient stability problems. 

\section*{Acknowledgements}
The work done by Alireza Soroudi is supported by a research grant from Science Foundation Ireland (SFI) under the SFI Strategic Partnership Programme Grant No. SFI/15/SPP/E3125. The opinions, findings and conclusions or recommendations expressed in this material are those of the author(s) and do not necessarily reflect the views of the Science Foundation Ireland. 
\newpage
\section*{References}

\bibliography{mybibfile}

\end{document}